\documentstyle[aps,epsf,prl,multicol]{revtex}

\begin{document}

\title{Fourier law in the alternate mass hard-core potential chain}

\author{
Baowen Li$^a$, Giulio Casati$^{b,c,a}$, Jiao Wang$^{d}$, and Toma\v z Prosen$^{e}$
}

\address{
$^a$Department of Physics, National University of Singapore,  Singapore 117542, Republic of Singapore\\
$^b$Center for Nonlinear and Complex Systems, 
Universita' degli studi dell'Insubria, Como\\
$^c$Istituto Nazionale di Fisica della Materia, Unita' di Como, and \\
Istituto Nazionale di Fisica Nucleare, sezione di Milano, 
Milano, Italy\\
$^d$ Temasek Laboratories, National University of Singapore, Singapore 119260, Republic of Singapore\\
$^e$ Physics Department, Faculty of Mathematics and Physics, University of 
Ljubljana, Ljubljana, Slovenia 
}
\date{\today}
\maketitle

\begin{abstract}

We study energy transport in a one-dimensional model of elastically colliding particles with alternate masses $m$ and $M$. In order to prevent total momentum conservation we confine particles with mass $M$ inside a cell of finite size. We provide convincing numerical evidence for the validity of Fourier law of heat conduction in spite of the lack of exponential dynamical instability. Comparison with previous results on similar models shows the relevance of the role played by total momentum conservation. 

\end{abstract}

%%%%%%%%%%%%%%%%%%%%%%%
\begin{multicols}{2}

After several decades of intensive investigations\cite{nmc,sum,alonso,triangle,Hatano,Campbell,2mass,narayan,LiWang}, the precise conditions that a dynamical 
system of interacting particles in 1D must satisfy in order to obey the Fourier law of heat conduction are still not known.

For non-interacting particles in external potential it has been shown \cite{alonso} that exponential local instability leads to Fourier law. 
Actually, even linear mixing without exponential instability, such as found in generic polygonal billiards\cite{CP},
has been shown to be sufficient for a diffusive heat transport\cite{triangle}.
In addition, several interacting non-integrable many-particle systems which clearly obey the Fourier law 
have been proposed and investigated\cite{nmc}.
However, it should be noted that in the above models, the total momentum is not conserved.
In several recent papers\cite{Hatano,Campbell,narayan} it has been suggested that total momentum conservation does not allow Fourier law.
Moreover,
using renormalization group\cite{narayan}, it is argued that a generic
momentum conserving particle chain should, in a macroscopic limit, be 
equivalent to 1D hydrodynamics with thermal noise where the coefficient of thermal conductivity should diverge with the system size $L$ as
$\kappa(L) \propto L^{1/3}$. However, most existing numerical data do not support this universal constant. Instead, it has been proposed that $\kappa(L) \propto L^{2-2/\alpha}$ \cite{LiWang}, where $\alpha$ is the exponent of the diffusion ($\langle \Delta x^2\rangle =2D t^{\alpha}$), $0 < \alpha \le 2$. 

We would like to remark that there exist a model\cite{giardina} (a particle chain with interparticle potential $V(x) = 1-\cos(x)$) in which, in spite of momentum conservation,
the heat conduction seems to obey the Fourier law. The reason for such behaviour is not clear and the precise role of the total momentum conservation needs to be clarified.

In previous papers two models have been considered, both are mixing and without exponential instability: 
(i) the triangular billiard channel \cite{triangle}, which exhibits Fourier law and
(ii) the alternate mass hard point gas
model\cite{2mass} in which the coefficient of thermal conductivity diverges with the system 
size. The difference between the two models is that in case (ii) the total momentum is conserved while in case (i) it is not. 

On the other hand, model (i) has been criticized since, in spite of the fact that 
one can perfectly well define an internal local temperature, there is no mechanism to provide local thermal equilibrium due to lack of interparticle interaction, and this may look somehow unsatisfactory.
In this paper we consider a model which is identical to the alternate mass hard-point gas \cite{2mass}, namely it consists of a 
one-dimensional chain of elastically colliding particles with alternate masses $m$ and $M$.
Here, however, in order
to prevent total momentum conservation we confine the motion of particles of mass $M$ (bars) inside unit cells of size $l$. 
Schematically the model is shown in Fig. 1 in which particles with mass $m$ move horizontally and collide with bars of 
mass $M$ which, besides suffering collisions with the particles, are elastically reflected back at the edges of their cells. 
In between collisions, particles and bars move freely.
Our numerical results clearly indicate that our model, contrary to the
translationally invariant model\cite{2mass}, obeys the Fourier law. 
The only difference between the two models is total momentum conservation. 

%%%%%%%%%%%%%Fig 1%%%%
\begin{figure}
\epsfxsize=8.cm \epsfbox{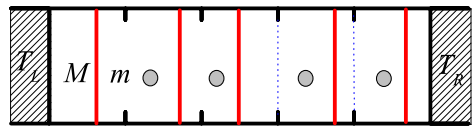}
\vspace{0cm}
\narrowtext
\caption {The geometry of the model.  The elementary cell (indicated
by two dotted lines) has unit length $l=1$. The bars have mass $M=1$, and
the particles have mass $m= (\sqrt{5}-1)/2$. The two heat
baths at temperatures $T_L$ and $T_R$ are indicated. }
\label{model}
\end{figure}
%%%%%%%%%%%%%%%%%%%%%%%%%

The total length of the system is $L= Nl$ where $N$ is the number of fundamental cells. In all the calculations presented
in this paper, we fix  $l=1$ so that $L=N$. We also take $M=1$, $m=(\sqrt{5}-1)/2$ and we verify that the numerical value of the mass ratio is not relevant.

A direct way to test whether the system obeys the Fourier law is to put two heat 
baths with small temperature difference into contact with the two ends of the system, 
and check the dependence of the thermal current on the system size. Here 
statistical thermal baths are used; that is when the first (last) bar collide
with the left (right) side of the first (last) cell, it is injected back with a new 
speed generated from the distribution
\begin{equation}
P_{L,R}(v) =
\frac{M|v|}{T_{L,R}}\exp\left(-\frac{Mv^2}{2T_{L,R}}\right).
\label{Gaussian}
\end{equation}
This assures, for small temperature gradients, that the edge particles have canonical 
(Maxwellian) velocity distribution.
In our simulations we fixed $T_L=1.1$ and $T_R=0.9$.

%%%%%%%%%%%%%Fig 2%%%%
\begin{figure}
\epsfxsize=8.cm \epsfbox{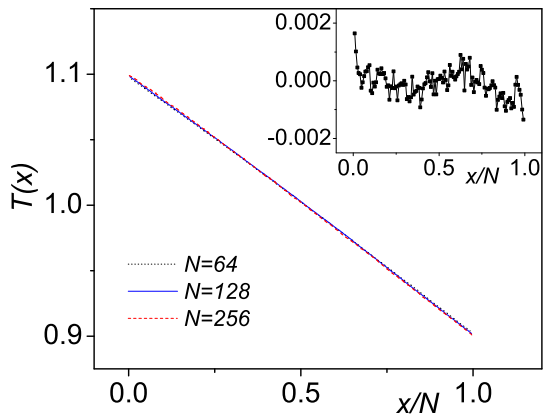}
 \vspace{0cm}
\narrowtext \caption{Internal local temperature as a function of the rescaled position
$x/N$. The total number $N$ of cells is: $N=64$ (dotted line), $N=128$ (solid line), and 
$N=256$ (dashed line). Notice the good scaling behavior of the temperature field.
The inset shows kurtosis of the local velocity distribution.}
\end{figure}
%%%%%%%%%%%%%%%%%%%%%%%%%

For any given initial condition, after a long enough transient time, the system 
reaches a stationary state.  Then one may compute the local temperature,
defined as
\begin{eqnarray}
T(x) &= &\langle Mv_i^2\rangle\quad \mbox {for}\quad
x=x_{i}^{\rm bar}=(i-0.5), \quad i=1,\cdots N\nonumber\\
T(x) &=&
 \langle mu_j^2\rangle\quad \mbox
{for}\quad x=x_{j}^{\rm part}=j,\quad j=1,\cdots N.
\end{eqnarray}
Here $x_{i}^{\rm bar}$ and $x_{j}^{\rm part}$  can be regarded roughly as the time-averaged 
positions of the $i$th bar and the $j$th particle. 

In Fig.2 we plot the temperature profile versus the scaled length $x/N$ for different
values of $N$. Notice the good linear scaling behaviour. In both Fig.2 and Fig.3, the average time measured by the total
collisions number is larger than $5\times 10^{10}$ for $N= 512$ which is the largest system size 
we have considered.

We should emphasize that in previous models\cite{alonso,triangle,2mass} the local thermal
equilibrium cannot be established, whereas in the model considered
here, the local thermal equilibrium is well established
independently of the thermal baths used. We have
checked that the velocity distribution function for each bar and
particle is a Gaussian function whose width gives the local
temperature, whereas the kurtosis, defined by $ K(x^{\rm bar}_i)=\frac{\langle v_i^4
\rangle}{3\langle v_i^2\rangle^2}-1,$ for bars and $K(x^{\rm part}_j)=\frac{\langle u_j^4 \rangle}{3\langle u_j^2\rangle^2}-1,$ for particles, are close to zero. The
kurtosis versus the bar/particle site is plotted in the inset of Fig.
2.

In Fig. 3 we show the stationary time-averaged heat flux $\langle j\rangle$ as a function of the system size $N$.  
The best fit of numerical data gives $\langle j\rangle=0.24N^{-\gamma}$ with $\gamma =0.99\pm 0.01$. The coefficient 
of thermal conductivity appears therefore to be independent on $N$, which means that the
Fourier law is obeyed. Its numerical value reads as $\kappa =-\frac{\langle j\rangle}{\nabla T}=1.20\pm 0.05$.

%%%%%%%%%%%%%Fig 3%%%%
\begin{figure}
\epsfxsize=8.cm \epsfbox{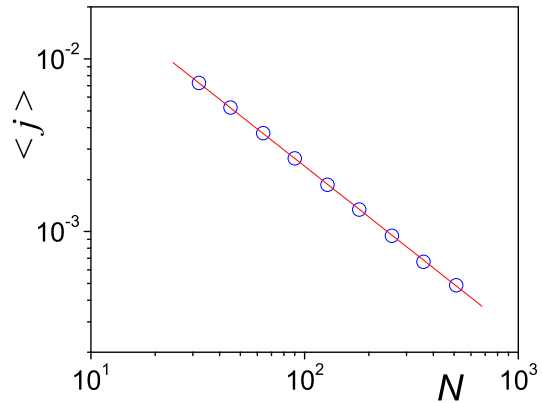} \vspace{0.3cm} \narrowtext
\caption{Scaling behavior of the stationary time-averaged heat flux $\langle j\rangle$ as a
function of the system size $N$. The least square fit gives a 
slope $-0.99 \pm 0.01$.}
\end{figure}
%%%%%%%%%%%%%%%%%%%%%%%%%

To investigate how the energy diffuses along the system, we set $T_L=T_R=1$. 
Then, after the equilibrium state is reached, 
the middle particle (i.e. the 100th one for $N=200$ in Fig. 4a) is given a speed $u=2/\sqrt{m}$, so that its energy is four times bigger than the equilibrium
average. The evolution of the energy profile along the chain is then recorded afterwards. To suppress statistical fluctuations,
$10^6$ realizations are taken into account for the average. The width of the energy profile can be measured by its second moment
\begin{equation}
\sigma^2(t)=\frac{\int(E(x,t)-E_0)(x-x_0)^2dx}{\int
(E(x,t)-E_0)dx},
\end{equation}
where $E(x,t) = \sum_j \frac{m u_j^2}{2} \delta(x-x^{\rm part}_j) +
\sum_i \frac{M v_i^2}{2}\delta(x-x^{\rm bar}_i)$.
In our calculations for Fig. 4, $E_0=0.5$ and $x_0=100$. The energy profile spreads as 
$\sigma^2(t) = 2Dt$ with $D=1.20\pm 0.01$ (Fig.4b), which agrees with the thermal conductivity
$\kappa=1.20\pm 0.05$ very well. 

%%%%%%%%%%%%%Fig 4a%%%%
\begin{figure}
\epsfxsize=8.cm \epsfbox{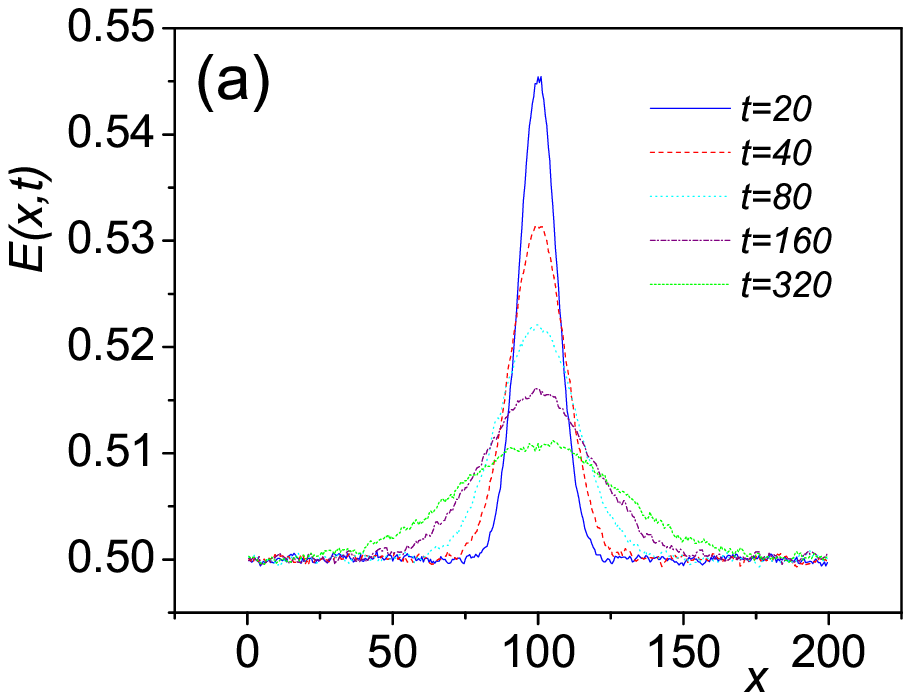}
\end{figure}
%%%%%%%%%%%%%%%%%%%%%%%%%

%%%%%%%%%%%%%Fig 4b%%%%
\begin{figure}
\vspace{-1cm}\epsfxsize=8.cm \epsfbox{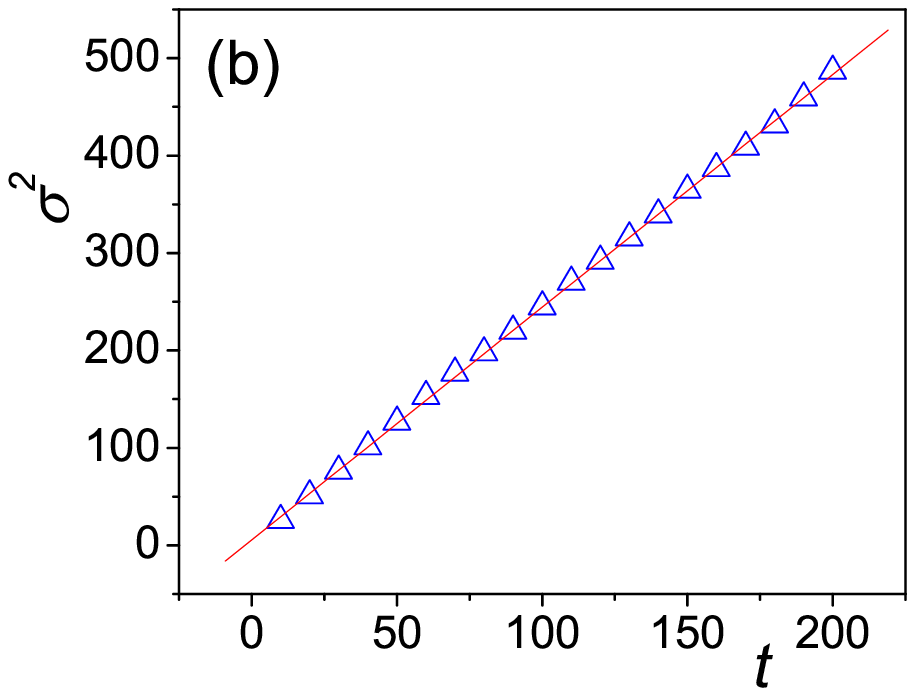} \narrowtext \caption
{Diffusive property of our model. Initially the system is at equilibrium with temperature $T=1$; 
then an impulse is given to the particle in the middle of 
the chain. (a) Energy distribution along the channel at different times. 
(b) The second moment of the energy distribution versus time. } \label{diffusion}
\end{figure}
%%%%%%%%%%%%%%%%%%%%%%%%%

If the system obeys the Fourier law, its thermal conductivity can also be obtained via
the Green-Kubo formula
\begin{equation}
\kappa=\lim_{t\to\infty}\lim_{N\to\infty}
\frac{1}{NT^2}\int_0^{t} \langle J(\tau)J(0)\rangle d\tau,
\end{equation}
where the heat current can be written as \cite{note}

\begin{equation}
J=\sum_{i=1}^N \frac{1}{2}\left(Mv_i^3+mu_i^3\right).
\end{equation}
In applying Green-Kubo theory, a periodic boundary condition is imposed to the system.

%%%%%%%%%%%%%Fig 5%%%
\begin{figure}
\epsfxsize=8.cm \epsfbox{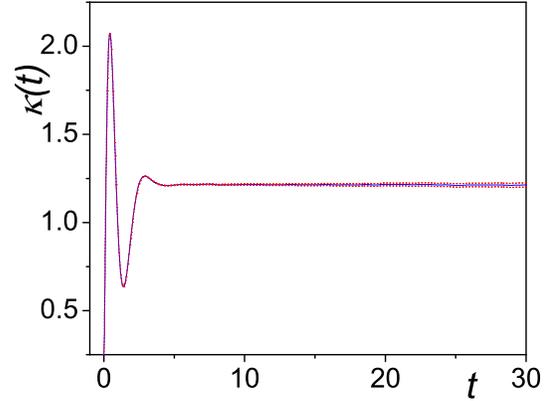} \narrowtext \caption {
The Green-Kubo integral versus $t$ for $N=24$ and $T=1$. To get $\langle J(\tau)J(0) \rangle$
used in (6), a single orbit is followed up to time $t_l=4.8\times 10^6$; the error, 
$\delta \langle J(\tau)J(0) \rangle$, is estimated in the same way as in Fig.6.
Dotted lines (almost indistinguishable) indicate the error boundaries of $\kappa(t)$, obtained by replacing the integrand in (6)
by $\langle J(\tau)J(0)\rangle\pm \delta \langle J(\tau)J(0)\rangle$ respectively. The errors are very small and negligible.
}
\label{Green-kubo1}
\end{figure}

In Fig. 5 we plot the quantity
\begin{equation}
\kappa(t)= \frac{1}{NT^2}\int_0^t \langle J(\tau)J(0)\rangle d\tau,
\end{equation}
as a function of time $t$ for $N=24$. Since $\langle J(\tau)J(0)\rangle$ decays
in time very fast (see Fig.6), one has that $\kappa(t)$ tends to $\kappa$ very fast 
as well. The numerical result gives $\kappa = 1.22$,
in excellent agreement to the heat conductivity obtained via simulations with thermal baths ($\kappa = 1.20$).

%%%%%%%%%%%%%Fig 6%%%

\begin{figure}
\epsfxsize=8.cm 
\epsfbox{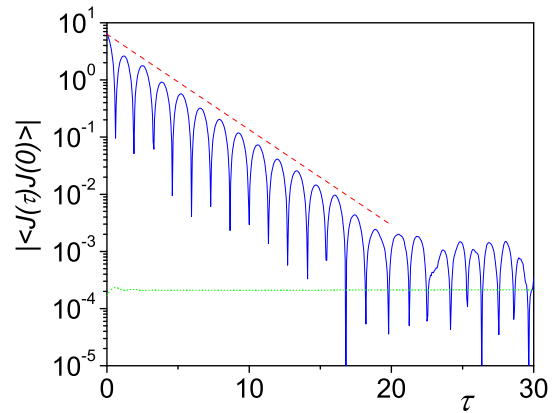} \narrowtext \caption {Absolute
heat current correlation function for $N=2$ and $T=1$. The time average is taken over a single orbit up to time $t_l=5\times10^7$. The dotted line 
shows the standard deviation of $\langle J(\tau)J(0) \rangle$ estimated over
$(t_l-\tau)/\delta\tau_0$ values of $J(\tau+\tau_0)J(\tau_0)$
obtained by changing $\tau_0$ with a step $\delta\tau_0=0.05$ 
along the orbit. The dashed line indicates an exponential decay 
$\exp(-0.38\tau)$.} 
\label{Green-kubo2}
\end{figure}
%%%%%%%%%%%%%%%%%%%%%%%%%

%%%%%%%%%%%%%%%%%%%%%%%%%

In Fig.6 we plot the absolute 
value of current-current time correlation function $C(\tau)=\langle J(\tau)J(0)\rangle$ 
versus $\tau$ for $N=2$. It is interesting to remark that numerical results seem to indicate a clear exponential decay
of correlation $C(\tau)$ which seems in contradiction with the linear, marginally unstable, dynamics. Indeed, as shown in \cite{CP}, the existence of periodic orbits in a marginally unstable system necessarily implies an asymptotic power-law decay of
correlation.
The asymptotic power-law tail may be however very difficult or impossible to observe numerically. In fact, what
we see is a transient exponential decay over several orders of magnitude which is very robust against changing
system parameters, such as $m/M$ or $N$.

Finally, we study the behaviour of the thermal conductivity $\kappa$ versus the mass ratio $m/M$ (Fig. 7). It is interesting to note that even at $m=M=1$ one finds a finite conductivity 
$\kappa$, which is close to the maximum of the curve in Fig. 7, in spite of the fact that
in this case ($m=M$) the dynamics is {\em pseudo-integrable} since, for
the isolated system, the set of magnitudes of 
the velocities of initial particles is conserved. However, the system
$m=M$ is {\em not strictly integrable} since the topology of invariant surfaces
is more complex than the one of the tori.
Notice also that only local thermal equilibrium is absent for $m=M$.

%%%%%%%%%%%%%Fig 7%%%
\begin{figure}
\epsfxsize=8.cm \epsfbox{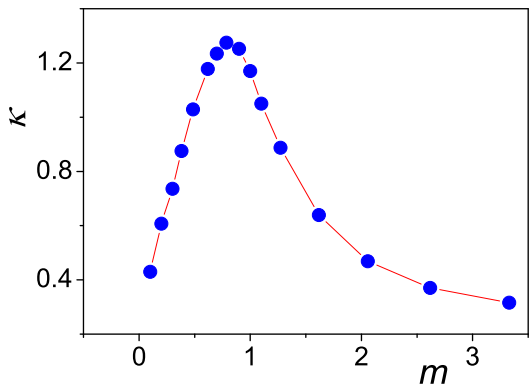} \narrowtext \caption {
The thermal conductivity $\kappa$ (bullet) versus the mass $m$ for $N=24$ and $T=1$. $M=1$. The solid curve is drawn to guide the eyes.
}
\label{Kappa-m}
\end{figure}
%%%%%%%%%%%%%%%%%%%%%%%%%

In the present paper we have demonstrated diffusive energy transport and Fourier law for
a marginally stable (non-chaotic) interacting many-particle system.
We have thus clearly demonstrated that exponential instability (Lyapunov chaos) is not necessary for 
the establishment of the Fourier law. Furthermore, we have shown that breaking the total momentum conservation is crucial for the validity of Fourier law while, somehow surprisingly, a less important role seems to be played by the degree of dynamical chaos.

\bigskip

BL is supported in part by Academic Research Fund of National University of Singapore. 
GC is partially supported by EU Contract 
No. HPRN-CT-2000-0156(QTRANS) and by MURST (Prin 2000, 
\emph{Caos e localizzazione in Meccanica Classica e Quantistica}).
JW is supported by the DSTA
of Singapore under Project Agreement POD0001821, and TP is supported
by the Ministry of Science, Education and Sport of Republic of Slovenia.

%\end{multicols}
%\end{document}

\end{multicols}

\begin{thebibliography}{9}
\bibitem{note} Notice that  an alternative
expression for the particle current in terms of energy transfer during collisions (such as that used in e.g.\cite{nmc}) seems to be more appropriate. The results however are the same.
\end{thebibliography}
\end{document}